\documentclass{sf2a-conf2013}
\usepackage{graphicx}
\usepackage{hyperref}
\usepackage[]{natbib}  
\usepackage{epstopdf}

\def\BibTeX{{\rm B\kern-.05em{\sc i\kern-.025em b}\kern-.08em
    T\kern-.1667em\lower.7ex\hbox{E}\kern-.125emX}}
\bibpunct{(}{)}{;}{a}{}{,}  

\newcommand{\ha}{\ifmmode {\rm H}\alpha \else H$\alpha$\fi}
\newcommand{\hb}{\ifmmode {\rm H}\beta \else H$\beta$\fi}
\newcommand{\lya}{\ifmmode {\rm Ly}\alpha \else Ly$\alpha$\fi}
\def\micron{$\mu$m}

\def\msun{\ifmmode M_{\odot} \else $M_{\odot}$\fi}
\def\msunyr{\ifmmode M_{\odot} {\rm yr}^{-1} \else $M_{\odot}$ yr$^{-1}$\fi}
\def\zsun{\ifmmode Z_{\odot} \else Z$_{\odot}$\fi}
\def\lsun{\ifmmode L_{\odot} \else L$_{\odot}$\fi}
\def\hii{{\rm HII}}
\def\mstar{M$_{\star}$}

\begin{document}

\TitreGlobal{SF2A 2013}


\title{Emission line properties from broad-band photometry: impact on selection and physical parameter estimation}

\runningtitle{Emission line properties from broad-band photometry}

\author{S. de Barros}\address{Department of Physics and Astronomy, University of California, Riverside, 900 University Avenue, Riverside, CA 92521, USA}

\author{H. Nayyeri$^{1}$}

\author{N. Reddy$^{1}$}

\author{B. Mobasher$^{1}$}

\setcounter{page}{237}


\maketitle


\begin{abstract}
Several works have now shown that nebular emission can have a significant impact on broad-band photometry of high-redshift galaxies ($z>3$), and how this can affect parameter estimation from SED fitting. So far relatively small spectroscopic samples have been used to measure this effect. Here we focus on a large spectroscopic sample (N$\sim2300$) at $z\sim2$, with measured H$\alpha$ fluxes for $\sim100$ galaxies. Under appropriate assumptions, our SED fitting code is able to reproduce observed \ha\ fluxes, and we infer that $\sim20$\% of our sample have parameters significantly affected when nebular emission is taken into account.

We also determine how nebular emission can affect selection and parameter estimation of evolved galaxies (Balmer Break galaxies) at high-redshift ($z\sim4$).\end{abstract}



\section{Introduction}
   Although nebular emission (i.e.\ emission lines and nebular continuous emission from \hii\ regions) is ubiquitous in regions of massive star formation, strong or dominant   in optical spectra of nearby star forming galaxies, and present in numerous types of galaxies, its impact on the determination of physical parameters of galaxies, in particular at high redshift, has been neglected until recently \citep[e.g.][]{zackrissonetal2008,schaerer&debarros2011}.
   
The analysis of samples of $z \sim$ 6--8 and $z\sim$ 3--6 LBGs observed with {\it HST} and {\it Spitzer} further demonstrates the potential impact of nebular emission on the physical parameters derived from SED fits of high-z galaxies \citep{schaerer&debarros2010,debarrosetal2012,schaereretal2013}. It has now become clear \citep{schaerer&debarros2009,schaerer&debarros2010,onoetal2010,lidmanetal2012} 
that nebular emission (both lines and continuum emission) must be taken
into account for the interpretation of photometric measurements of the SEDs of star-forming galaxies at high-z. In parallel, diverse evidence of galaxies with strong emission lines and/or strong contributions 
of nebular emission to broad-band fluxes has been found at different redshifts, e.g.\ by
\cite{shimetal2011,mclindenetal2011,ateketal2011,trumpetal2011,vanderweletal2011,labbeetal2012,starketal2013,smitetal2013}. Unfortunately, at r$z>3$, direct measurement of emission lines is still challenging with current facilities, although some studies at $3.0<z<3.8$ show promising results \citep[e.g.][]{schenkeretal2013}.

The main idea of this work is to use a spectroscopic sample at $z\sim2$ \citep{erbetal2006}, where measurements of emission lines (\ha) are already available for a significant subsample. This subsample is used to test the ability of our SED fitting code (but the results can be generalized to any SED fitting code accounting for nebular emission) to reproduce properly observed emission lines, under different assumptions. After this calibration, we infer the impact of nebular emission on parameter estimation. Furthermore, we also determine how nebular emission can affect selection and physical parameter estimation of evolved galaxies at $z\sim4$ \citep{nayyerietal2013}.
We adopt a $\Lambda$-CDM cosmological model with $H_{0}$=70 km s$^{-1}$ Mpc$^{-1}$, $\Omega_{m}$=0.3 and $\Omega_{\Lambda}$=0.7. 
All magnitudes are expressed in the AB system \citep{oke&gunn1983}.

\section{SED fitting code}
We use a recent, modified version of the Hyperz photometric redshift code of \cite{bolzonellaetal2000}, taking into account nebular emission (lines and continua). We consider a large set of spectral templates \citep{bruzual&charlot2003},
covering different metallicities and a wide range of star formation (SF) histories (exponentially decreasing, constant and exponentially rising SF). Nebular emission from continuum processes and lines is added to the spectra predicted from the GALAXEV models as described in 
\cite{schaerer&debarros2009}, proportionally to the Lyman continuum photon production. The relative line intensities of He
 and metals are taken from \cite{anders&fritz2003}, including galaxies grouped in three metallicity intervals covering 
 $\sim$ 1/50--1  \zsun. Hydrogen lines from the Lyman to the Brackett series are included with relative intensities given by case B.
We adopt a spectral templates computed for a Salpeter IMF \citep{salpeter1955} from 0.1 to 100 $M_\odot$, and we 
properly treat the returned ISM mass from stars. The IGM is treated following \cite{madau1995}. For galactic attenuation we use the Calzetti law \citep{calzettietal2000}. With these assumptions
we fit the observed SEDs by straightforward least-square minimization.

We test two attenuations, one applying the same attenuation to the stellar and nebular emission, and the other following \cite{calzetti1997}, who find a more important attenuation for nebular emission in comparison with stellar emission, with $E(B-V)_{\star}=0.44\times E(B-V)_{neb}$.

\section{Data}

Galaxies at redshifts $1.4\leq z \leq3.7$ were selected using the BM, BX, and Lyman-break galaxy (LBG) rest-UV color criteria \citep{steideletal2003,steideletal2004,adelbergeretal2004}. The imaging data were obtained mostly with the Palomar Large Format Camera or the Keck Low Resolution Imaging Spectrograph \citep{okeetal1995,steideletal2004}. The photometry and spectroscopic follow-up for this survey are described in \cite{steideletal2003,steideletal2004,adelbergeretal2004}. To probe the strength of the Balmer break in $z\sim2$ galaxies, we used J and/or Ks imaging (Palomar/WIRC and Magellan/PANIC). We also used Spitzer/IRAC data. The IRAC coverage of our galaxies typically included either channels 1 (3.6\micron) and 3 (5.8\micron)
or channels 2 (4.5\micron) and 4 (8.0\micron), with a small fraction of galaxies having coverage in all four channels. Additionally, for a small fraction of this $z\sim2$ sample, we obtained (HST)/WFC3-F160W (H band) data.

The search and characterization of evolved galaxies at $z\sim4$ has been done with Wide Field Camera 3 (WFC3) near Infra Red observations in the GOODS-S field performed as a part of the CANDELS project. The U-band data in the GOODS-S area were taken using the VIMOS instrument on Very Large Telescope. Optical data comes from HST/ACS, with F435W, F606W, F775W and F850LP filters. Near-infrared observations are in the F098M and F105W filters with the former one being used for the GOODS-S ERS, F125W and F160W. GOODS-S has also observations by the VLT HAWK-I Ks filter at effective wavelength of 2.2 \micron. Finally, we also use Spitzer/IRAC data in all four channels \citep[for a detailed description of the data, see][and references therein]{nayyerietal2013}. In IRAC Ch1 and Ch2, there is the much deeper data as part of the SEDs program through Spitzer warm mission.



\section{Physical parameters of $z\sim2$ star-forming galaxies}
\begin{figure}[ht!]
 \centering
 \includegraphics[width=8.25cm,trim=0.75cm 0.25cm 1.5cm 1.25cm,clip=true]{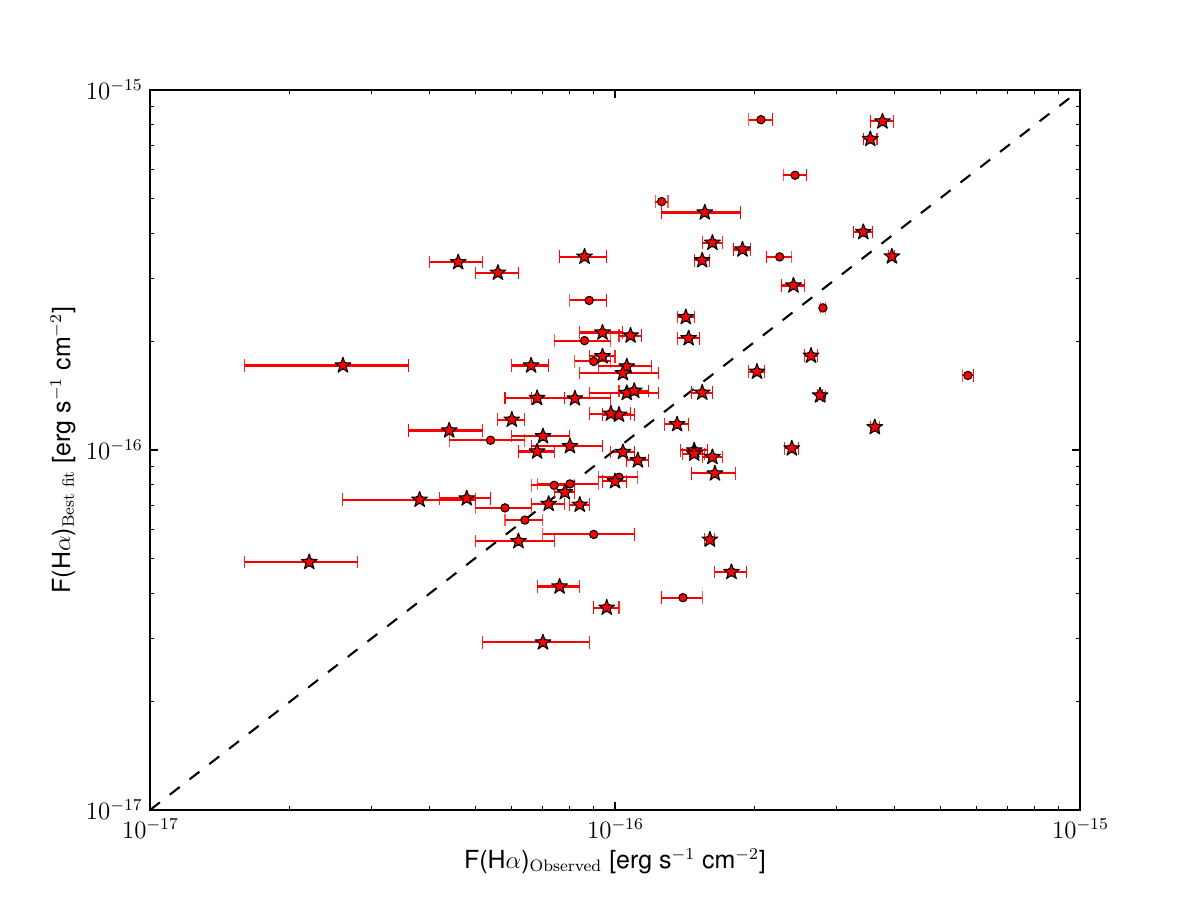}      
  \includegraphics[width=8.25cm,trim=0.75cm 0.25cm 1.5cm 1.25cm,clip=true]{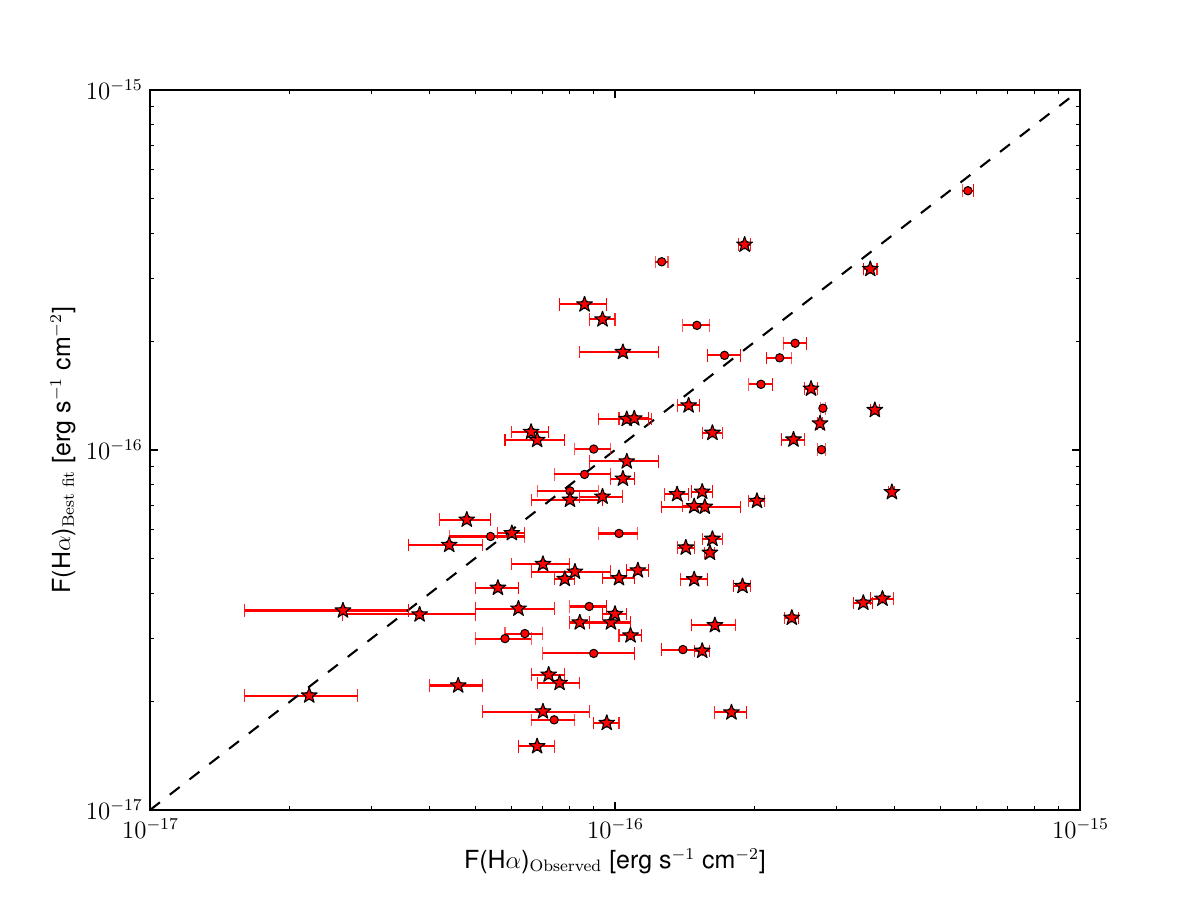}      
  \caption{{\bf Left:} \ha\ fluxes inferred from best-fit SED fitting vs observed \ha\ fluxes, with a rising SFH, applying the same attenuation to the stellar and nebular emission. {\bf Right:} same but with a differential attenuation \citep[][]{calzetti1997}. Red stars: galaxies with \ha\ falling in one filter (F160 or K band), red dots: galaxies with \ha\ line falling in no filter.}
  \label{debarros:fig1}
\end{figure}

First, we test the ability of the three star formation histories (exponentially declining/rising, and constant) used here, to reproduce observed \ha\ fluxes, since these different histories lead to different predictions of line strength \citep{debarrosetal2012,schaereretal2013}. Only declining SFH is unable to reproduce observed fluxes, leading to lower f(\ha)$_{\mathrm{SED}}$ in comparison with f(\ha)$_{\mathrm{obs}}$. In Figure \ref{debarros:fig1}, we show the result with a rising SFH (similar results with a constant SFH). In the local Universe, \cite{calzetti1997} show the existence of a differential attenuation between nebular and stellar emission. This is explained by nebular and stellar emission are not emitted from the same regions (nebular emission comes from star-forming regions), and that dust properties are not the same in these regions. Applying the same differential attenuation than \cite{calzetti1997} leads overall to underpredict \ha\ fluxes (Figure \ref{debarros:fig1}, right). However, for some individual objects, this leads to improve the correlation between observed and predicted \ha\ fluxes. Similarly, \cite{kashinoetal2013} find that the \cite{calzetti1997} relation evolves at higher redshift, which may indicate a more uniform dust distribution in high-z galaxies as compared to local galaxies.

We also test SED fitting with a lower age limit ($>50$ Myr), to avoid conflict between dynamical timescale and age. Using this limit  marginally improve the agreement between predicted and observed fluxes.

Relying on our consistency test between predicted and observed  \ha\ flux, we only use the rising SFH (since constant SF leads to similar results) SFH, with same attenuation between stellar and nebular emission to infer physical properties of our $z\sim2$ sample, and determine how nebular emission affects parameter estimation. The main parameter that we expect to be affected by nebular emission is age \citep{schaerer&debarros2009}, since some lines (H$\beta$, [OIII] doublet, [OII]) can mimic a Balmer break. We compare results of our fit without and with nebular emission, and we find that 468 galaxies (20\% of our sample) have $\log(\mathrm{Age})-\log(\mathrm{Age_{NEB}})\ge0.3$.

\begin{figure}[ht!]
 \centering
 \includegraphics[width=8.25cm,trim=0.75cm 0.25cm 1.5cm 1.25cm,clip=true]{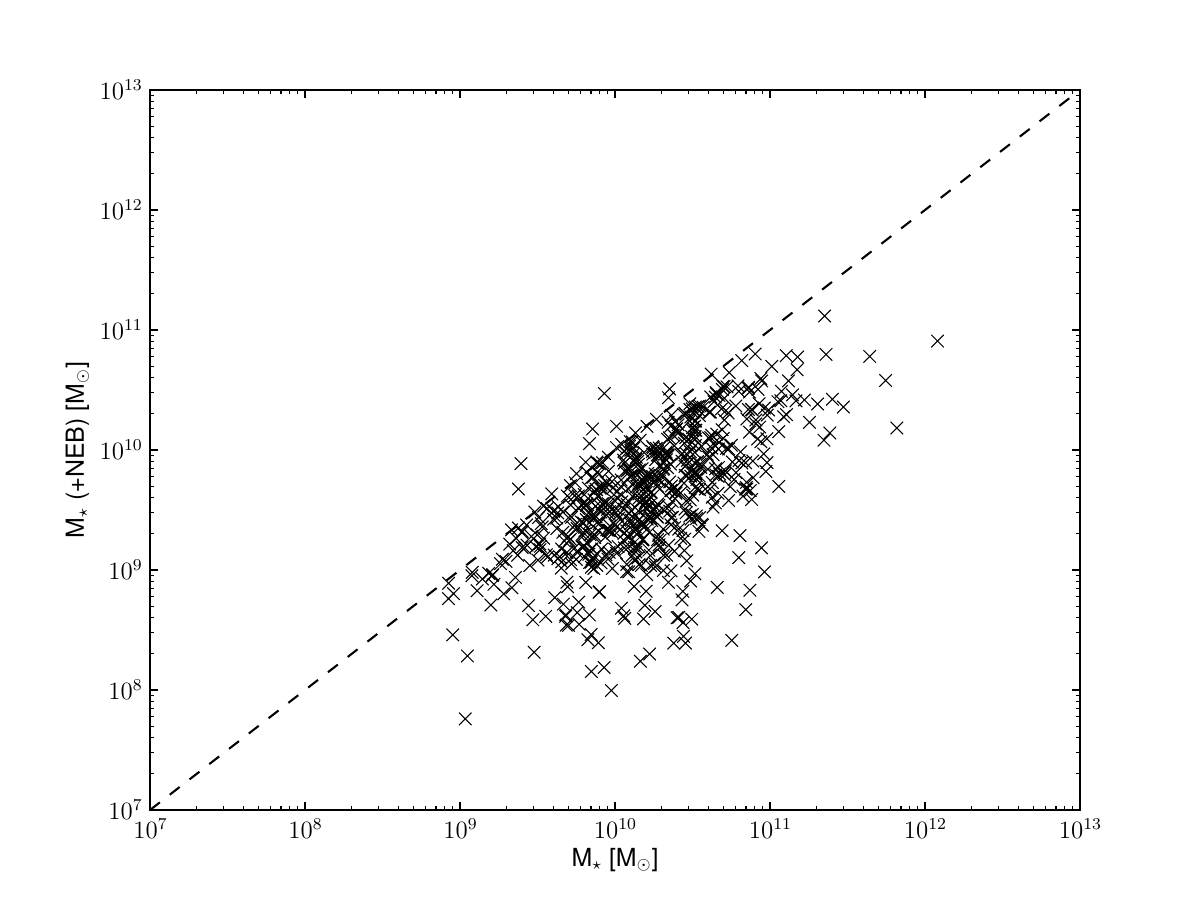}      
  \caption{Stellar mass estimated without nebular emission vs stellar mass estimated with nebular emission for the whole spectroscopic sample ($1.41<z<3.57$), and with $\log(\mathrm{Age})-\log(\mathrm{Age_{NEB}})\ge0.3$. The dashed is the one to one relation.}
  \label{debarros:fig2}
\end{figure}

At $z\sim2$, stellar masses are strongly constrained by IRAC data, since no strong emission lines fall in these filters. However, the stellar mass estimation can be affected by nebular emission if age is significantly affected too, since the mass to light ratio evolves with age. Age estimation can be modified if [OII] and/or [OIII] lines affect a band that lies close to the Balmer break. Indeed, in Figure \ref{debarros:fig2}, we show the comparison between stellar mass estimated without nebular emission (\mstar) and stellar mass estimated with nebular emission (\mstar$_{\mathrm{NEB}}$), for galaxies for which age estimation changes when accounting for nebular emission ($\log(\mathrm{Age})-\log(\mathrm{Age_{NEB}})\ge0.3$). The impact of nebular emission is however not very strong since we find only 293 galaxies (12\% of the sample) with $\log(\mathrm{M_{\star}})-\log(\mathrm{M_{\star NEB}})\ge0.3$

In another paper, we will present detailed results and implications for star formation history.

\begin{figure}[ht!]
 \centering
 \includegraphics[width=8.25cm,trim=0.75cm 0.25cm 1.5cm 1.25cm,clip=true]{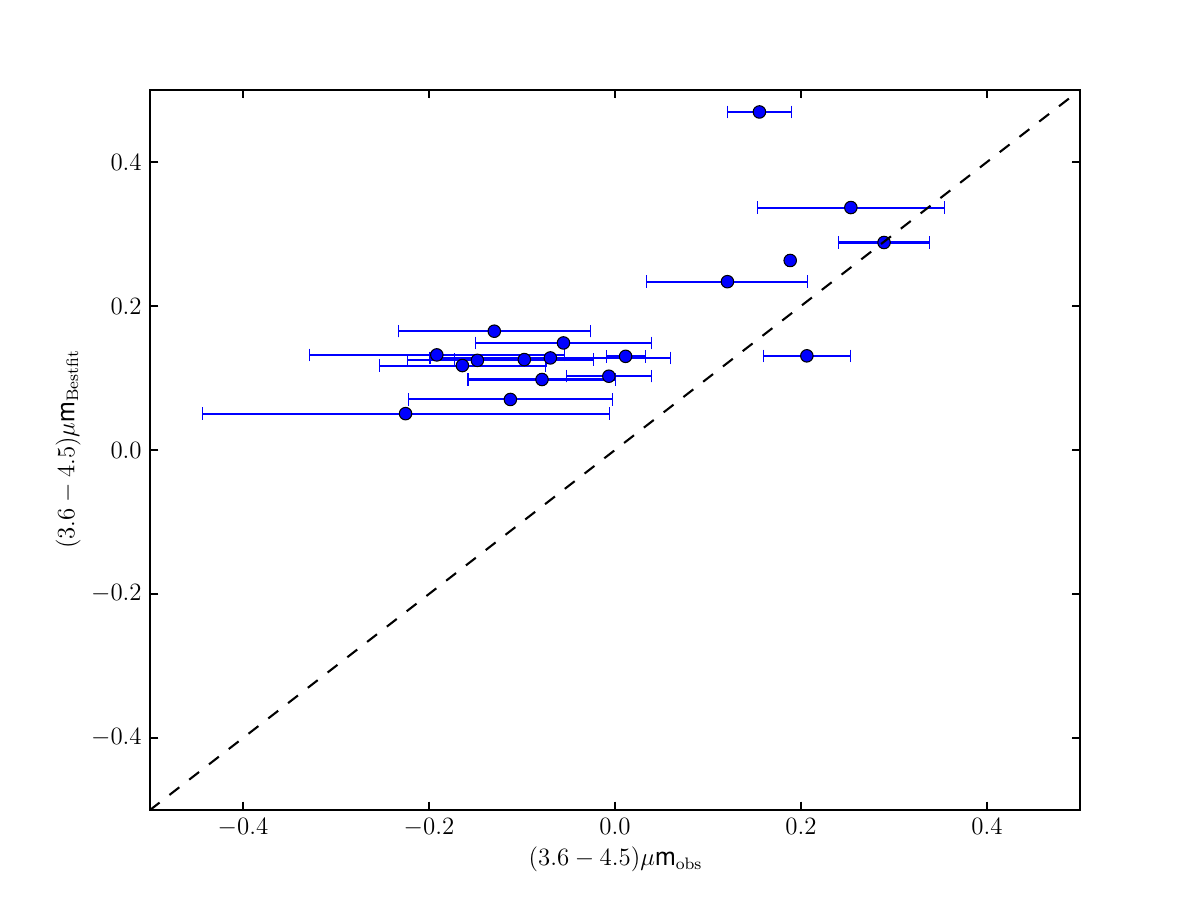}      
  \includegraphics[width=8.25cm,trim=0.75cm 0.25cm 1.5cm 1.25cm,clip=true]{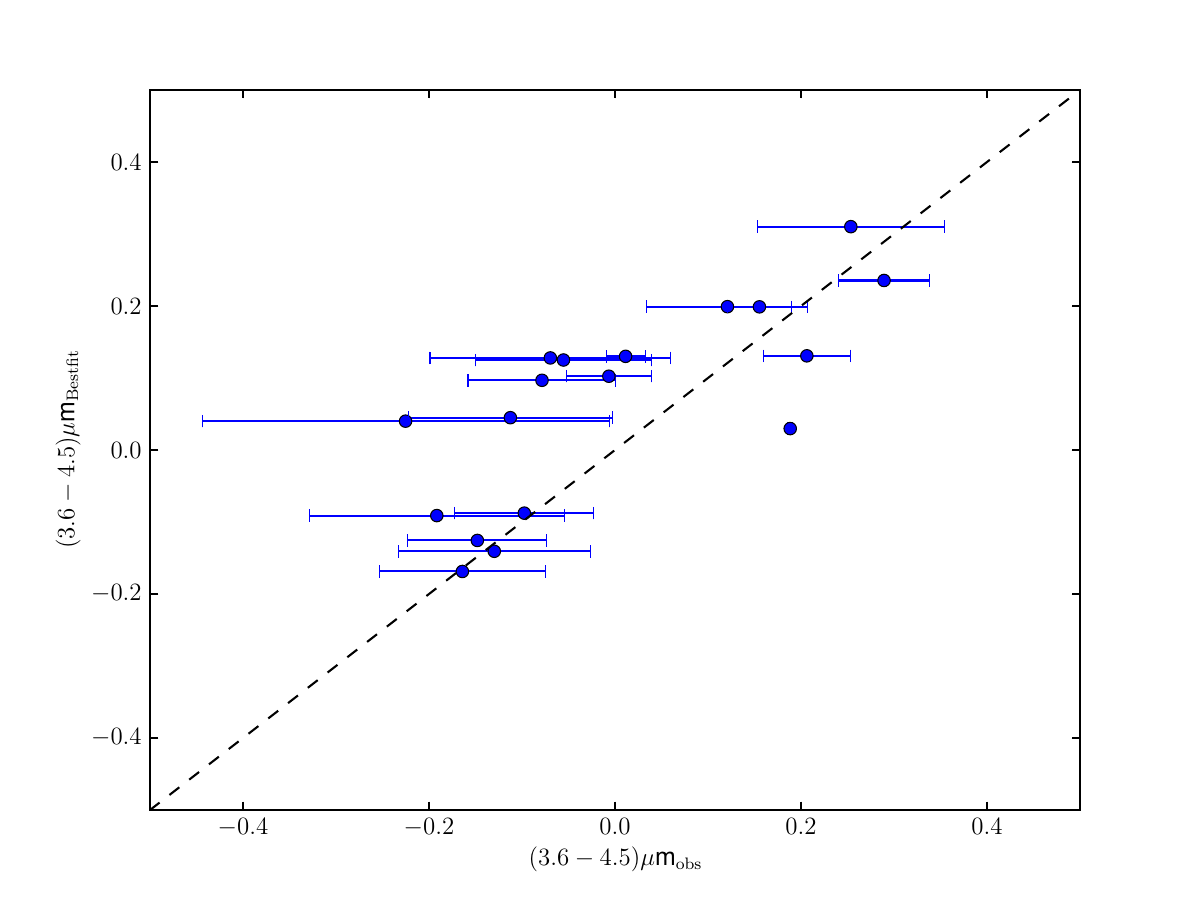}      
  \caption{{\bf Left:} (3.6-4.5)\micron\ observed color vs (3.6-4.5)\micron\ best-fit SED color for objects with $3.8<z_{\mathrm{phot}}<5.0$, without taking into account nebular emission. {\bf Right:} same but accounting for nebular emission.}
  \label{debarros:fig3}
\end{figure}

\section{Selection and physical properties of evolved galaxies at $z\sim4$}

Using evolutionary tracks, we define a color-color selection to identify evolved galaxies at $z\sim4$, i.e\ with a strong Balmer break \citep[see][for selection criteria]{nayyerietal2013}.
Contamination is mainly due to dusty starburst galaxies, while nebular emission can allow low-redshift starburst to pass the selection in narrow ranges of redshift, when flux in selection filters is enhanced by some emission lines.

The physical properties of $z\sim4$ galaxies are more strongly affected by nebular emission in comparison with $z\sim2$ galaxies, since strong emission lines (e.g. \ha) affect IRAC fluxes at $z>3.8$. While no current facilities allow direct measurements of emission lines (except \lya) at those redshift, \ha\ line is found in the 3.6\micron\ filter at $3.8<z<5.0$, whereas very few lines are expected in the 4.5\micron\ filter. Excess in the 3.6\micron\ filter is interpreted as an additional flux due to the \ha\ line, and the (3.6-4.5)\micron\ color is basically a model independent measurement of the \ha\ strength \citep{shimetal2011,debarrosetal2012,starketal2013}. In Figure \ref{debarros:fig3}, we show the ability of our SED fitting code to reproduce observed (3.6-4.5)\micron\ color of objects identified as evolved galaxies. As at $z\sim2$, we show that our SED fitting code provide fits consistent with the observed strength of \ha.

SED fitting of our complete sample of evolved galaxies shows that an upper limit of $\sim20$\% of the sample show contamination by emission lines, which affects age, stellar mass, dust and star formation rate estimation. Typically, ages are reduced, stellar masses decreased, and dust attenuation and SFR increased. The complete analysis of these evolved galaxies can be found in \citep{nayyerietal2013}.

\section{Conclusions}

We present a homogeneous study of a sample of 2389 spectroscopically confirmed star-forming galaxies at $z\sim2$, with deep photometry up to 8 $\mu$m. We also use available measurement of \ha\ line for $\sim100$ galaxies. Using a modified version of the {\em HyperZ} photometric redshift code which takes into account nebular emission \citep{schaerer&debarros2009}, we explore a range of star formation histories (constant, exponentially decreasing and rising). We test different star formation histories and differential attenuation between stellar and nebular emission \citep{calzetti1997} to reproduce observed \ha\ fluxes. Declining SFH is unable to provide consistent f(\ha)$_{\mathrm{SED}}$, while differential attenuation leads also to underestimated f(\ha)$_{\mathrm{SED}}$ in comparison with observed fluxes. The main parameter affected by nebular emission is age, since some lines can mimic a Balmer break \citep{schaerer&debarros2009,schaerer&debarros2010}. While stellar mass is strongly constrained by IRAC data, changes in age estimation also affects significantly stellar mass estimation for a fraction of our sample (12\%).

At $z\sim4$, evolved galaxies are more affected when nebular emission is accounted for in SED fitting, since strong emission lines can affect IRAC fluxes. Selection of such galaxies is mostly contaminated by dusty starburst galaxies, with only little effect of nebular emission. While we do not have access to observational evidence of strong emission lines at this redshift, the (3.6-4.5)\micron color provide empirical evidence of strong \ha\ emission, color reproduced consistently by our SED fitting tool. $\sim20$\% of our sample identified as evolved galaxies at $z\sim4$ significantly affect by nebular emission, leading to younger ages, lower stellar masses, higher dust attenuation and higher SFRs. Complete results of this study are presented in \cite{nayyerietal2013}.

\begin{acknowledgements}
We thank the SF2A organisers. SdB also thanks the SF2A for its financial support. SdB is supported by a Swiss National Science Foundation Early.Postdoc fellowship.
\end{acknowledgements}


\bibliographystyle{aa} 
\bibliography{ref}

\end{document}